\documentstyle[12pt]{article}

\begin{document}
\title{Coefficients of Chiral Perturbation Theory}
\author{Bing An Li\\
Department of Physics and Astronomy, University of Kentucky\\
Lexington, KY 40506, USA }

\maketitle

\begin{abstract}
Based on an effective chiral theory of pseudoscalar, vector, and
axial-vector mesons, the coefficients of the chiral perturbation
theory are predicted. There is no new parameter in these predictions.
\end{abstract}

\newpage
The chiral perturbation theory(ChPT)
proposed a decade ago[1] is successful
in parametrizing low-energy $QCD$. The chiral symmetry revealed from
$QCD$, quark mass expansion, and momentum expansion
are used to construct the Lagrangian of ChPT
\begin{eqnarray}
\lefteqn{{\cal L}={f^{2}_{\pi}\over16}TrD_{\mu}UD^{\mu}U^{\dag}+
{f^{2}_{\pi}\over16}Tr\chi(U+U^{\dag})+
L_{1}[Tr(D_{\mu}UD^{\mu}U^{\dag})]^{2}+L_{2}(TrD_{\mu}UD_{\nu}U^
{\dag})^{2}}\nonumber \\
&&+L_{3}Tr(D_{\mu}UD^{\mu}U^{\dag})^{2}+L_{4}Tr(D_{\mu}U
D^{\mu}U^{\dag})Tr\chi(U+U^{\dag})
+L_{5}TrD_{\mu}UD^{\mu}U^{\dag}(\chi U^{\dag}+U\chi)\nonumber \\
&&+L_{6}[Tr\chi(U+U^{\dag})]
^{2}+L_{7}[Tr\chi(U-U^{\dag})]^{2}+L_{8}
Tr(\chi U\chi U+\chi U^{\dag}\chi U^{\dag})
\nonumber \\
&&-iL_{9}Tr(F^{L}_{\mu\nu}D^{\mu}UD^{\nu}U^{\dag}+F^{R}_{\mu\nu}
D^{\mu}U^{\dag}D^{\nu}U)+L_{10}Tr(F^{L}_{\mu\nu}UF^{\mu\nu R}U^{\dag}).
\end{eqnarray}
The parameters
in the chiral perturbation theory(1) are pion decay constant
$f_{\pi}$ and the 10 coefficients. These parameters are determined
by fitting experimental data[1,2,3,4,5,6,7].

The chiral perturbation theory is rigorous and phenomenologically
successful in describing the physics of the pseudoscalar mesons at low
energies.
Models attempt to deal with the two main frustrations that
the ChPT is limited to pseudoscalar mesons at
low energy and contains many coupling
constants which must be measured. However, the chiral perturbation
theory sets a low energy limit for all models.
The price paid for including more
mesons and determining the coefficients is
to make additional assumption. Many models try to
predict the chiral couplings(see Table 2).
In Ref.[8] an extended Nambu-Jona-Lasinio model(ENJL)
\[{\cal L}=\bar{q}\{\gamma^{\mu}(v_{\mu}+\gamma_{5}a_{\mu})-
(s-ip\gamma_{5})\}q\]
has been used to predict the coefficients of the ChPT.
Nine of the 10 coefficients, $L_{1,2,3,4,5,6,8,9,10}$,
are predicted in Ref.[8b] and listed in Table II.

I have proposed an effective chiral theory of pseudoscalar, vector, and
axial-vector mesons[9,10]. It provides a unified description
of low-lying meson physics. This effective theory is phenomenologically
successful. It is natural to study whether the Lagrangian(1)
of the chiral perturbation theory can be derived from this
effective theory and the 10 coefficients can be predicted.
The Lagrangian of this theory is
\begin{eqnarray}
{\cal L}=\bar{\psi}(x)(i\gamma\cdot\partial+\gamma\cdot v
+\gamma\cdot a\gamma_{5}
-mu(x))\psi(x)-\bar{\psi(x)}M\psi(x)\nonumber \\
+{1\over 2}m^{2}_{0}(\rho^{\mu}_{i}\rho_{\mu i}+K^{*\mu}K^{*}_{\mu}+
\omega^{\mu}\omega_{\mu}+\phi^{\mu}\phi_{\mu}
+a^{\mu}_{i}a_{\mu i}+K^{\mu}_{1}K_{1\mu}
+f^{\mu}f_{\mu}+f^{\mu}_{1s}f_{1s\mu})
\end{eqnarray}
where M is the quark mass matrix
\[\left(\begin{array}{c}
         m_{u}\hspace{0.5cm}0\hspace{0.5cm}0\\
         0\hspace{0.5cm}m_{d}\hspace{0.5cm}0\\
         0\hspace{0.5cm}0\hspace{0.5cm}m_{s}
        \end{array}  \right ),\]
\(v_{\mu}=\tau_{i}\rho^{i}_{\mu}+\lambda_{a}K^{*a}_{\mu}
+({2\over3}+{1\over\sqrt{3}}\lambda_{8})\omega_{\mu}
+({1\over3}-{1\over\sqrt{3}}\lambda_{8})\phi_{\mu}\),
\(a_{\mu}=\tau_{i}a^{i}_{\mu}+\lambda_{a}K^{a}_{1\mu}
+({2\over3}+{1\over\sqrt{3}}\lambda_{8})f_{\mu}+({1\over3}
-{1\over\sqrt{3}}\lambda_{8})f_{1s\mu}\),
and \(u=exp\{i\gamma_{5}(\tau_{i}\pi_{i}+\lambda_{a}K_{a}+\lambda_{8}
\eta_{8}+\eta_{0})\}\).
$u$ can be written as
\begin{equation}
u={1\over 2}(1+\gamma_{5})U+{1\over 2}(1+\gamma_{5})U^{\dag},
\end{equation}
where \(U=exp\{i(\tau_{i}\pi_{i}+\lambda_{a}K_{a}+\lambda_{8}
\eta+\eta_{0})\}\).

In Ref.[11] this Lagrangian has been applied to study $\tau$ mesonic
decays systematically. Theoretical results agree well with data.
In this paper the Lagrangian(2) is used to predict all the 10 coefficients
of ChPT.

As a matter of fact, in Ref.[9] two of the
coefficients have been determined from the contribution of the $\rho$
resonance in
$\pi\pi$ scattering. There are contact terms which make less contribution.
In this paper a complete expressions of the coefficients $L_{1,2,3}$
are presented.

The vertex $\rho\pi\pi$ is[9]
\begin{eqnarray}	
\lefteqn{{\cal L}^{\rho\pi\pi}=f_{\rho\pi\pi}(q^{2})
\epsilon_{ijk}\rho^{i}_{\mu}
\pi_{j}\partial^{\mu}\pi_{k},}\nonumber \\
&&f_{\rho\pi\pi}(q^{2})={2\over g}\{1+\frac{q^{2}}{2\pi^{2}f^{2}_{\pi}}
[(1-{2c\over g})^{2}-4\pi^{2}c^{2}]\},\nonumber \\
&&c=\frac{f^{2}_{\pi}}{2gm^{2}_{\rho}},
\end{eqnarray}
where g is the universal coupling constant in this effective theory[9] and
q is the momentum of the $\rho$ meson.
The effective Lagragian of $\pi\pi$ scattering at low energy
($q^{2}<m^{2}_{\rho}$) is derived from Eq.(4) and Eq.(13) of Ref.[9]
\begin{eqnarray}
\lefteqn{{\cal L}={1\over2f^{2}_{\pi}}\partial_{\mu}\pi^{2}
\partial^{\mu}\pi^{2}}\nonumber \\
&&+{16\over f^{4}_{\pi}}\{[{1\over2}{1\over(4\pi)^{2}}(1-{2c\over g})^{2}
(-1+{4c\over g}+{12c^{2}\over g^{2}})-{c^{4}\over g^{2}}]\partial_{\mu}
\pi_{i}\partial^{\mu}\pi_{i}\partial_{\nu}\pi_{j}\partial^{\nu}\pi_{j}
\nonumber \\
&&+[{1\over(4\pi)^{2}}(1-{2c\over g})^{2}(1-{4c\over g}-{4c^{2}\over g^{2}
})+{c^{4}\over g^{2}}]
\partial_{\mu}
\pi_{i}\partial^{\nu}\pi_{i}\partial_{\mu}\pi_{j}\partial^{\nu}\pi_{j}\}
\nonumber \\
&&-{4\over f^{4}_{\pi}}(1-{2c\over g}){2c\over g}\{2gc+{1\over\pi^{2}}(
1-{2c\over g})\}\{
\partial_{\mu}
\pi_{i}\partial^{\mu}\pi_{i}\partial_{\nu}\pi_{j}\partial^{\nu}\pi_{j}-
\partial_{\mu}
\pi_{i}\partial^{\nu}\pi_{i}\partial_{\mu}\pi_{j}\partial^{\nu}\pi_{j}\}.
\end{eqnarray}
Comparing with Lagrangian of the ChPT(1), we obtain
\begin{eqnarray}
\lefteqn{2(L_{1}+L_{2})+L_{3}={1\over4}{1\over(4\pi)^{2}}(1-{2c\over g})^{4}}
\nonumber \\
&&L_{2}={1\over4}{c^{4}\over g^{2}}+{1\over4}{1\over(4\pi)^{2}}(1-{2c\over
g})^{2}(1-{4c\over g}-{4c^{2}\over g^{2}})\nonumber \\
&&+{1\over8}(1-{2c\over g}){c\over g}\{2gc+{1\over\pi^{2}}(1-{2c\over g})\}.
\end{eqnarray}
Only two of the three coefficients are determined from $\pi\pi$ scattering.
Therefore, another process
is needed. We choose $\pi$K scattering[12] at low energy to do the job.
$\rho$ and $K^{*}$ resonances contribute to $\pi$K scattering. Using Eq.(13)
in Ref.[9], the $\rho\pi\pi$ vertex and
\begin{equation}
{\cal L}^{K^{*}\pi\pi}=f_{\rho\pi\pi}(q^{2})f_{abk}K^{a*}_{\mu}(K_{b}\partial
^{\mu}\pi_{k}-\pi_{k}\partial^{\mu}K_{b})[11],	
\end{equation}
in the chiral limit the two isospin amplitudes of $\pi$K scattering
at low energy are
derived as
\begin{eqnarray}
\lefteqn{T^{{3\over2}}=-{2\over f^{2}_{\pi}}s}\nonumber \\	
&&+{16\over f^{4}_{\pi}}\{{1\over(4\pi)^{2}}{4c^{2}\over g^{2}}(1-{2c\over g}
)^{2}-{c^{4}\over 2g^{2}}\}\{t^{2}+u^{2}\}\nonumber \\
&&+{16\over f^{4}_{\pi}}\{{1\over(4\pi)^{2}}(1-{2c\over g}
)^{2}(1-{4c\over g}-{4c^{2}\over g^{2}})+{c^{4}\over g^{2}}\}
s^{2}\nonumber \\
&&-{1\over f^{4}_{\pi}}{4c\over g}(1-{2c\over g})\{2gc+{1\over \pi}
(1-{2c\over g})\}\{u(s-t)+t(s-u)\},\nonumber \\
&&T^{{1\over2}}=-{1\over f^{2}_{\pi}}(3u-s)\nonumber \\
&&+{16\over f^{4}_{\pi}}\{{1\over(4\pi)^{2}}{4c^{2}\over g^{2}}(1-{2c\over g}
)^{2}-{c^{4}\over2g^{2}}\}t^{2}\nonumber \\
&&-{8\over f^{4}_{\pi}}\{{1\over(4\pi)^{2}}(1-{2c\over g}
)^{2}(1-{4c\over g}-{16c^{2}\over g^{2}})+{5\over2}
{c^{4}\over g^{2}}\}s^{2}\nonumber \\
&&+{8\over f^{4}_{\pi}}\{{1\over(4\pi)^{2}}(1-{2c\over g}
)^{2}(3-{12c\over g}-{16c^{2}\over g^{2}})+{7\over2}
{c^{4}\over g^{2}}\}u^{2}\nonumber \\
&&+{1\over f^{4}_{\pi}}{2c\over g}(1-{2c\over g})\{2gc+{1\over\pi}
(1-{2c\over g})\}\{-3s(u-t)+u(s-t)+4t(s-u)\},
\end{eqnarray}
where
\[s=(p_{1}+p_{2})^{2},\;\;\;t=(p_{1}-p_{3})^{2},\;\;\;u=(p_{1}-p_{4})^{2}
,\]
$p_{1,3}$ are the momenta of initial and final pions and $p_{2,4}$
are the momenta of initial and final kaons respectively.
The two isospin amplitudes of $\pi$K scattering can be derived from the
chiral perturbation theory(1) too. Comparing the partial waves obtained from
Eqs.(8) and ChPT,
besides Eqs.(6) obtained from $\pi\pi$ scattering we obtain
\begin{eqnarray}
\lefteqn{-2L_{1}+L_{2}=0,}\nonumber \\
&&L_{3}=-{3\over16}{2c\over g}(1-{2c\over g})\{2gc+{1\over\pi^{2}}
(1-{2c\over g})\}
-{1\over2}{1\over(4\pi)^{2}}(1-{2c\over g})^{2}(1-{4c\over g}-{8c^{2}\over
g^{2}})-{3\over4}{c^{4}\over g^{2}},\nonumber \\
&&L_{1}={1\over32}{2c\over g}(1-{2c\over g})\{2gc+{1\over\pi^{2}}
(1-{2c\over g})\}
+{1\over8}{1\over(4\pi)^{2}}(1-{2c\over g})^{2}(1-{4c\over g}
-{4c^{2}\over g^{2}
})+{1\over4}{c^{4}\over g^{2}}.
\end{eqnarray}
The first equation of Eq.(9) has been obtained in Ref.[8].

It is learned from Ref.[9] that
$g^{2}\sim O(N_{C})$, $f^{2}_{\pi}\sim
O(N_{C})$, and $m^{2}_{\rho}\sim O(1)$ in large $N_{C}$ expansion.
Therefore, $L_{1,2,3}\sim O(N_{C})$.
The predictions of $L_{1-3}$ are made at the tree level, therefore,
the Eq.(9) actually means that $-2L_{1}+L_{2}\sim O(1)$ in
large $N_{C}$ expansion.

According to Ref.1(b), the coefficients $L_{4-8}$ are determined from
the quark mass expansions of
$m^{2}_{\pi}$, $m^{2}_{K}$, $m^{2}_{\eta}$, $f_{\pi}$, $f_{K}$, and
$f_{\eta}$.
Therefore, the quark mass term of the Lagrangian(5) is needed to be
taken into account in
deriving the effective Lagrangian of mesons.
The
expressions of the masses and the
decay constants of the pseudoscalars
are found from the real part of the effective
Lagrangian. Using the method presented in Ref.[9],
in Euclidean space the real part of the effective Lagrangian of mesons
with
quark masses is written as
\begin{eqnarray}
\lefteqn{{\cal L}_{RE}={1\over2}\int d^{D}x\frac{d^{D}p}{(2\pi)^{D}}
\sum^{\infty}_{n=1}{1\over n}\frac{1}{(p^{2}+m^{2})^{n}}}
\nonumber \\
&&Tr\{(\gamma\cdot\partial-i\gamma\cdot v+i\gamma\cdot a\gamma_{5})
(\gamma\cdot\partial-i\gamma\cdot v-i\gamma\cdot a\gamma_{5})
+2ip\cdot(\partial-iv-ia)\nonumber \\
&&+m\gamma\cdot Du
-i[\gamma\cdot v,M]+i\{\gamma\cdot a,M\}\gamma_{5}
-m(\hat{u}M+Mu)-M^{2}\}^{n}.
\end{eqnarray}
where \(D_{\mu}u=\partial_{\mu}u-i[v_{\mu},u]+i\{a_{\mu},u\}
\gamma_{5}\) and \(\hat{u}=
exp\{-i\gamma_{5}[\tau_{i}\pi_{i}+\lambda_{a}
K_{a}+\lambda_{8}\eta_{8}+\eta_{0}]\}\).
Comparing with Eq.(11) of Ref.[9],
there are new terms in which the
quark mass matrix M is involved.
As done in Ref.[9], the mesons fields in Eq.(10) have to be
normalized.

The masses of the pseudoscalcar mesons are derived from Eq.(10).
The masses of the pseudoscalars
can be calculated
to any order in quark masses.
Up to the second order in quark masses we obtain
\begin{eqnarray}
\lefteqn{m^{2}_{\pi^{\pm}}={4\over f^{2}_{\pi^{\pm}}}\{-{1\over3}
<\bar{\psi}\psi>(m_{u}+m_{d})-{F^{2}\over4}(m_{u}+m_{d})^{2}
\},}\nonumber \\
&&m^{2}_{\pi^{0}}={4\over f^{2}_{\pi^{0}}}\{-{1\over3}
<\bar{\psi}\psi>(m_{u}+m_{d})-{F^{2}\over2}(m^{2}_{u}+m^{2}_{d})
\},\nonumber \\
&&m^{2}_{K^{+}}={4\over f^{2}_{K^{+}}}\{-{1\over3}
<\bar{\psi}\psi>(m_{u}+m_{s})-{F^{2}\over4}(m_{u}+m_{s})^{2}
\},\nonumber \\
&&m^{2}_{K^{0}}={4\over f^{2}_{K^{0}}}\{-{1\over3}
<\bar{\psi}\psi>(m_{d}+m_{s})-{F^{2}\over4}(m_{d}+m_{s})^{2}
\},\nonumber \\
&&m^{2}_{\eta_{8}}={4\over f^{2}_{\eta_{8}}}\{-{1\over3}
<\bar{\psi}\psi>{1\over3}(m_{u}+m_{d}+4m_{s})
-{F^{2}\over6}(m^{2}_{u}+m^{2}_{d}+4m^{2}_{s})\},
\end{eqnarray}
where $F^{2}$ is defined in Ref.[9] as
\[F^{2}={N_{C}\over \pi^{4}}m^{2}\int\frac{d^{4}p}
{(p^{2}+m^{2})^{2}},\]
$<\bar{\psi}\psi>$ is the quark condensate of three flavors.
In terms of a cut-off $\Lambda$ the quark condensate
is expressed as(using Eqs.(40) and (43) of Ref.[9])
\begin{equation}
<\bar{\psi}\psi>={i\over(2\pi)^{4}}
Tr\int d^{4}p\frac{\gamma\cdot p-mu}{p^{2}
-m^{2}}
=\frac{m^{3}N_{C}}{4\pi^{2}}
\{{\Lambda^{2}\over m^{2}}-log({\Lambda^{2}\over m^{2}}+1)\}.
\end{equation}
The universal coupling constant g is expressed as[9]
\begin{equation}
g^{2}={F^{2}\over6m^{2}}={1\over2\pi^{2}}\{log({\Lambda^{2}\over m^{2}}
+1)+\frac{1}{{\Lambda^{2}\over m^{2}}+1}-1\}.
\end{equation}
The decay constants of the pseudoscalars, $f_{\pi}$, $f_{K}$, and
$f_{\eta}$, are defined by normalizing the
pseudoscalar fields as done in Refs.[9,10].
In this paper
we calculate the contributions of the quark masses to these
decay constants to $O(m_{q})$.
Using the
effective Lagrangian(10), the decay constants can be calculated
to any order
in quark masses. As indicated in Refs.[9,10],
there is mixing between the
axial-vector field and corresponding pseudoscalar field. The mixing
results in the shifting of the axial-vector field $a_{\mu}$
\begin{equation}
a_{\mu}\rightarrow {1\over g_{a}}a_{\mu}
-{c_{a}\over g_{a}}\partial_{\mu}P,
\end{equation}
where P is the corresponding pseudoscalar field, $g_{a}$ is the
normalization constant of the $a_{\mu}$ field, and $c_{a}$
is the mixing
coefficient. Both $g_{a}$ and $c_{a}$ are determined
in the chiral limit in Ref.[9].
For pion and $a_{1}$ fields
it is obtained by eliminating the mixing between pion
and $a_{1}$ fields
\begin{equation}
{c_{a}\over g_{a}}={1\over g^{2}_{a}m^{2}_{a}}\{{F^{2}\over2}+
({F^{2}\over8m^{2}}-{3\over4\pi^{2}})2m(m_{u}+m_{d})\},
\end{equation}
where $m_{a}$ is the mass of the $a_{1}$ meson, which is determined from
the Lagrangian(10) as
\begin{equation}
g^{2}_{a}m^{2}_{a}=F^{2}+g^{2}m^{2}_{\rho}+6g^{2}_{0a}m(m_{u}+m_{d}),
\end{equation}
where $g^{2}_{0a}$ is expressed as[9]
\begin{equation}
g^{2}_{0a}=g^{2}(1-{1\over2\pi^{2}g^{2}}).
\end{equation}
Up to the first order in quark masses,
the decay constant $f^{2}_{\pi}$ is obtained from the Lagrangian(10)
\begin{equation}
f^{2}_{\pi}=f^{2}_{\pi0}\{1+f\frac{m_{u}+m_{d}}{m}\},
\end{equation}
where
\begin{equation}
f^{2}_{\pi0}=F^{2}(1-{2c\over g})[9]
\end{equation}
and
\begin{equation}
f=(1-{2c\over g})(1-
{1\over2\pi^{2}g^{2}})-1+\frac{4}{\pi^{2}f^{4}_{\pi0}}(-{1\over3})
<\bar{\psi}\psi>m(1-{2c\over g})(1-{c\over g}).
\end{equation}
In the same way, $f^{2}_{K^{+}}$, $f^{2}_{K^{0}}$, and
$f^{2}_{\eta_{8}}$ are found
\begin{eqnarray}
f^{2}_{K^{+}}=f^{2}_{\pi0}\{1+f\frac{m_{u}+m_{s}}{m}\},\\
f^{2}_{K^{0}}=f^{2}_{\pi0}\{1+f\frac{m_{d}+m_{s}}{m}\}, \\
f^{2}_{\eta_{8}}=f^{2}_{\pi0}\{1+{1\over3}f\frac{m_{u}+
m_{d}+4m_{s}}{m}\}.
\end{eqnarray}

Substituting Eqs.(18,21,22,23)
into Eq.(11), to the second order in quark masses
the masses of the pseudoscalar mesons are obtained
\begin{eqnarray}
\lefteqn{m^{2}_{\pi^{\pm}}={4\over f^{2}_{\pi0}}\{-{1\over3}
<\bar{\psi}\psi>(m_{u}+m_{d})-{F^{2}\over4}(m_{u}+m_{d})^{2}
+{f\over3}<\bar{\psi}\psi>{1\over m}(m_{u}+m_{d})^{2}
\},}\nonumber \\
&&m^{2}_{\pi^{0}}={4\over f^{2}_{\pi0}}\{-{1\over3}
<\bar{\psi}\psi>(m_{u}+m_{d})-{F^{2}\over2}(m^{2}_{u}+m^{2}_{d})
+{f\over3}<\bar{\psi}\psi>{1\over m}(m_{u}+m_{d})^{2}
\},\nonumber \\
&&m^{2}_{K^{+}}={4\over f^{2}_{\pi0}}\{-{1\over3}
<\bar{\psi}\psi>(m_{u}+m_{s})-{F^{2}\over4}(m_{u}+m_{s})^{2}
+{f\over3}<\bar{\psi}\psi>{1\over m}(m_{u}+m_{s})^{2}
\},\nonumber \\
&&m^{2}_{K^{0}}={4\over f^{2}_{\pi0}}\{-{1\over3}
<\bar{\psi}\psi>(m_{d}+m_{s})-{F^{2}\over4}(m_{d}+m_{s})^{2}
+{f\over3}<\bar{\psi}\psi>{1\over m}(m_{d}+m_{s})^{2}
\},\nonumber \\
&&m^{2}_{\eta_{8}}={4\over f^{2}_{\pi0}}\{-{1\over3}
<\bar{\psi}\psi>{1\over3}(m_{u}+
m_{d}+4m_{s})-{F^{2}\over4}{2\over3}(m^{2}_{u}+m^{2}_{d}
+4m_{s}^{2})\nonumber \\
&&+{f\over3}<\bar{\psi}\psi>{1\over9}{1\over m}
(m_{u}+m_{d}+4m_{s})^{2}\}\nonumber \\
&&={4\over f^{2}_{\pi0}}\{-{1\over3}
<\bar{\psi}\psi>{1\over3}(m_{u}+
m_{d}+4m_{s})
+({f\over m}{1\over3}<\bar{\psi}\psi>-{F^{2}\over 4})
{1\over9}
(m_{u}+m_{d}+4m_{s})^{2}\nonumber \\
&&-{2F^{2}\over9}[{1\over2}(m_{u}+m_{d})-m_{s}]^{2}
-{F^{2}\over12}(m_{d}-m_{u})^{2}\}.
\end{eqnarray}
The mass formulas(24) are the same as the ones
presented in Ref.(1(b)).
The mass difference of charged and neutral pions is found from
Eqs.(24)
\begin{equation}
m^{2}_{\pi^{\pm}}-m^{2}_{\pi^{0}}=(1-{2c\over g})^{-1}(m_{d}
-m_{u})^{2}.
\end{equation}
Comparing with Eqs.(24) with the Eqs.(10.7,10.8) of Ref.(1(b)),
following
coefficients of the ChPT are predicted
\begin{eqnarray}
L_{4}=0,\;\;\;\;L_{6}=0,\\
L_{5}=\frac{f^{2}_{\pi0}f}{8mB_{0}},\\
L_{8}=-\frac{F^{2}}{16B_{0}^{2}},\\
3L_{7}+L_{8}=-\frac{F^{2}}{16B_{0}^{2}},\;\;\;L_{7}=0,
\end{eqnarray}
where
\begin{equation}
B_{0}={4\over f^{2}_{\pi0}}(-{1\over3})<\bar{\psi}\psi>.
\end{equation}
Using Eq.(30), $L_{5}$ and $L_{8}$ are written as
\begin{eqnarray}
L_{5}=\frac{1}{32Q}(1-{2c\over g})\{(1-{2c\over g})^{2}
(1-{1\over2\pi^{2}g^{2}})-(1-{2c\over g})+{4\over \pi^{2}}Q
(1-{c\over g})\},\\
L_{8}=-\frac{1}{1536g^{2}Q^{2}}(1-{2c\over g})^{2},
\end{eqnarray}
where
\begin{equation}
Q=-{1\over108g^{4}}{1\over m^{3}}<\bar{\psi}\psi>.
\end{equation}
The Eqs.(12,13) show that Q is a function of the universal coupling
constant g only. By fitting $\rho\rightarrow e^{+}e^{-}$, g is determined
to be 0.39[11].
The numerical value of Q is 4.54.

Both $L_{5}$ and $L_{8}$ are at $O(N_{C})$. $L_{4}$, $L_{6}$, and
$L_{7}$ are from loop diagrams of mesons and are at $O(1)$ in large
$N_{C}$ expansion.

According to Refs.[1(b),3], $L_{9}$ and $L_{10}$ are determined by
$<r^{2}>_{\pi}$ and the amplitudes of pion radiative decay,
$\pi^{-}\rightarrow e^{-}\gamma\nu$
\begin{eqnarray}
L_{9}={f^{2}_{\pi}\over48}<r^{2}_{\pi}>,\\
L_{9}={1\over32\pi^{2}}{R\over F^{V}}, \\
L_{10}={1\over32\pi^{2}}{F^{A}\over F^{V}}-L_{9},
\end{eqnarray}
where R, $F^{V}$, and $F^{A}$
are $r_{A}$, $h_{V}$,and $h_{A}$ of Ref.[3] respectively.
In Ref.[11] we have derived the expression of pion radius as
\begin{equation}
<r^{2}>_{\pi}={6\over m^{2}_{\rho}}+{3\over\pi^{2}f^{2}_{\pi}}
\{(1-{2c\over g})^{2}-4\pi^{2}c^{2}\}
\end{equation}
which agrees with the data very well.
Using Eqs.(34,37), it is predicted
\begin{equation}
L_{9}={1\over4}cg+{1\over16\pi^{2}}
\{(1-{2c\over g})^{2}-4\pi^{2}c^{2}\}[14],
\end{equation}
On the other hand, $L_{9}$ is determined by
Eq.(35) too. The three form factors of the decay amplitude of
$\pi^{-}\rightarrow e^{-}\gamma\nu$ are presented in our paper[11]
\begin{eqnarray}
F^{V}=\frac{m_{\pi}}{2\sqrt{2}\pi^{2}f_{\pi}},\\
F^{A}={1\over2\sqrt{2}\pi^{2}}{m_{\pi}\over f_{\pi}}{m^{2}_{\rho}
\over m^{2}_{a}}(1-{2c\over g})(1-{1\over2\pi^{2}g^{2}})^{-1},
\\
R={g^{2}\over\sqrt{2}}{m_{\pi}\over f_{\pi}}{m^{2}_{\rho}\over
m^{2}_{a}}\{{2c\over g}+{1\over\pi^{2}g^{2}}(1-{2c\over g})\}
(1-{1\over2\pi^{2}g^{2}})^{-1}+\sqrt{2}cg{m_{\pi}\over f_{\pi}}
[15].
\end{eqnarray}
Using the mass formula of the $a_{1}$ meson(in the chiral limit)[9]
\[(1-{1\over2\pi^{2}g^{2}})m^{2}_{a}={F^{2}\over g^{2}}+m^{2}_
{\rho},\]
it is obtained
\begin{equation}
{m^{2}_{\rho}\over m^{2}_{a}}(1-{1\over2\pi^{2}g^{2}})^{-1}
=1-{2c\over g}.
\end{equation}
Substituting Eq.(42) into R(41), it is derived
\begin{equation}
R={1\over3\sqrt{2}}m_{\pi}f_{\pi}<r^{2}>_{\pi}.
\end{equation}
This is the expression obtained by applying PCAC to this process[14].
Therefore, the coefficient $L_{9}$ derived from Eq.(35) is the same
as the one obtained from Eq.(34).

Using Eq.(42), the ratio ${F^{A}\over F^{V}}$ is written as
\begin{equation}
{F^{A}\over F^{V}}=(1-{2c\over g})^{2}.
\end{equation}
The coefficient $L_{10}$ is found from Eq.(36)
\begin{equation}
L_{10}=-{1\over4}cg+{1\over4}c^{2}-{1\over32\pi^{2}}
(1-{2c\over g})^{2}.
\end{equation}
Both $L_{9}$ and $L_{10}$ are at $O(N_{C})$.

In the expressions of the coefficients(6,9,26,29,31,32,38,45)
there are two
parameters: g and $f^{2}_{\pi}/m^{2}_{\rho}$ which have been determined
already.
The coefficients of the ChPT are completely determined by
this effective chiral theory. The numerical values of the 10 coefficients
predicted by present theory are listed in table 3.

To summarize the results. It is necessary to point out that theoretical
results of
the processes, $\pi\pi$ scattering, $\pi$K[16] scattering, masses of the
pseudoscalar mesons, the radius of pion, and the form factors of
$\pi\rightarrow\gamma e\nu$ have been tested.
The orders of the coefficients of the ChPT
in large $N_{C}$ expansion are predicted as $L_{1,2,3,5,8,9,10}\sim
O(N_{C})$ and $L_{4,6}$ are $O(1)$. These predictions agree with
Ref.[1(b)]. It is also predicted that $L_{7}\sim O(1)$ in this paper.
It is necessary to point out that the expressions of the coefficients
$L_{4,5,6,7,8}$ are derived from Eq.(24), however, Eq.(11) is used to
fit the masses of the pseudoscalar mesons. It is found
that this way
leads to consistent determination of the quark masses.
The results will be presented somewhere else.
There are no new parameters in
predicting the coefficients of the ChPT.

The author wishes to thank B.Holstein for discussion.
This research was partially
supported by DOE Grant No. DE-91ER75661.

\begin{table}[h]
\begin{center}
\caption{Values of the coefficients[1(b)]}
\begin{tabular}{|c|c|c|c|c|c|c|c|c|c|} \hline
$10^{3}L_{1}$&$10^{3}L_{2}$&$10^{3}L_{3}$
&$10^{3}L_{4}$&$10^{3}L_{5}$&$10^{3}L_{6}$
&$10^{3}L_{7}$&$10^{3}L_{8}$&$10^{3}L_{9}$
&$10^{3}L_{10}$  \\ \hline
$.9\pm.3$&$1.7\pm.7$&-$4.4\pm 2.5$&$0.\pm.5$&
$2.2\pm.5$&$0.\pm0.3$&-$.4\pm.15$& $1.1\pm.3$&
$7.4\pm.7$&-$6.\pm.7$ \\ \hline
\end{tabular}
\end{center}
\end{table}
\begin{table}[h]
\begin{center}
\caption{Coefficients obtained by models}
\begin{tabular}{|c|c|c|c|c|c|c|c|} \hline
&Vectors[4]&Quark[2]&Ref.[6]&Ref.[7]&Nucleon loop&Linear$\sigma$
model&ENJL\\ \hline
$(L_{1}$+${1\over2}L_{3}$)$\times10^{-3}$&-2.1&-.8&2.1&1.1&-.8&-.5&  \\ \hline
$L_{1}\times10^{-3}$ & & & & & & &0.8                               \\ \hline
$L_{2}\times10^{-3}$ &2.1&1.6&1.6&1.8&.8&1.5&1.6  \\ \hline
$L_{3}\times10^{-3}$ & & & & & & &-4.1 \\ \hline
$L_{4}\times10^{-3}$ & & & & & & &0. \\ \hline
$L_{5}\times10^{-3}$ & & & & & & &1.5 \\ \hline
$L_{6}\times10^{-3}$ & & & & & & &0. \\ \hline
$L_{7}\times10^{-3}$ & & & & & & & \\ \hline
$L_{8}\times10^{-3}$ & & & & & & &0.8 \\ \hline
$L_{9}\times10^{-3}$ &7.3&6.3&6.7&6.1&3.3&.9&6.7 \\ \hline
$L_{10}\times10^{-3}$ &-5.8&-3.2&-5.8&-5.2&-1.7&-2.0&-5.5  \\ \hline
\end{tabular}
\end{center}
\end{table}
\begin{table}[h]
\begin{center}
\caption{Predictions of the Values of the coefficients}
\begin{tabular}{|c|c|c|c|c|c|c|c|c|c|}  \hline
$10^{3}L_{1}$&$10^{3}L_{2}$&$10^{3}L_{3}$
&$10^{3}L_{4}$&$10^{3}L_{5}$&$10^{3}L_{6}$
&$10^{3}L_{7}$&$10^{3}L_{8}$&$10^{3}L_{9}$
&$10^{3}L_{10}$  \\ \hline
1.0&2.0&-5.16&0&4.77&0&0&-0.079&8.3&-7.1\\ \hline
\end{tabular}
\end{center}
\end{table}


\begin{thebibliography}{40}
\bibitem{} In Eq.(67) of Ref.[19] the factor ${1\over2}$ should be
taken off and there should be an
additional term from the form factor
\(\frac{m^{2}_{\rho}}{m^{2}_{\rho}-k^{2}}\).
\end{thebibliography}
\end{document}